\documentclass[preprint,12pt]{elsarticle}

\usepackage{amssymb}
\usepackage{amsthm}
\usepackage{txfonts}

\newtheorem{thm}{Theorem}
\newtheorem{lemma}[thm]{Lemma}
\newtheorem{proposition}[thm]{Proposition}
\newtheorem{rem}{Remark}

\def\C{\mathbb{C}}

\def\N{\mathbb{N}}
\def\K{\mathbb{K}}

\def\gotM{\mathfrak{M}}
\def\gotMp{\mathfrak{M}^{+}}
\def\Kxp{\mathbb{K}[[\mathtt{x}]]^{+}}

\def\im{\mathsf{im}}
\def\End{\mathsf{End}}

\def\Id{\mathsf{Id}}
\def\UP{\mathsf{UM}}
\def\US{\mathsf{US}}

\def\HWS{{\cal H}_{WS}}
\def\1H{\mathbf{1}_{\HWS}}

\def\x{\mathtt{x}}
\def\mup{\mu_{+}}
\def\sigmap{\sigma_{+}}

\def\2m#1#2{(#2 #1)}
\def\3m#1#2#3{(#3 #2 #1)}

\begin{document}

\begin{frontmatter}



\title{A formal calculus on the Riordan near algebra}


\author[lipn]{L.~Poinsot}
\ead{laurent.poinsot@lipn.univ-paris13.fr}
\author[lipn]{G.H.E.~Duchamp}
\ead{ghed@lipn-univ.paris13.fr}
\address[lipn]{LIPN - UMR 7030\linebreak
CNRS - Universit\'e Paris 13\linebreak F-93430 Villetaneuse,
France}
\begin{abstract}
The Riordan group is the semi-direct product of a multiplicative group of invertible series and a group, under substitution,  of non units. The Riordan near algebra, as introduced in this paper, is the Cartesian product of the algebra of formal power series and its principal ideal of non units, equipped with a product that extends the multiplication of the Riordan group. The later is naturally embedded as a subgroup of units into the former. In this paper, we prove the existence of a formal calculus on the Riordan algebra. This formal calculus plays a role similar to those of holomorphic calculi in the Banach or Fr\'echet algebras setting, but without the constraint of a radius of convergence. Using this calculus, we define \emph{en passant} a notion of generalized powers in the Riordan group. 
\end{abstract}


\end{frontmatter}



\section{Introduction}
\label{Intro}

As defined in \cite{SGW$^+$91} a Riordan matrix is an infinite matrix $M^{(\mu,\sigma)}$$=$$(m_{i,j})_{i,j\geq 0}$ with complex coefficients such that for every $j\in\N$, the ordinary generating function of its $j$th column is equal to $\mu(\x)\sigma(\x)^j$, or in other terms,
for every $j\in \N$, $\displaystyle \sum_{i\geq 0}m_{i,j}\x^i = \mu(\x)\sigma(\x)^j$, where $\mu,\sigma$ are two formal power series in the variable $\x$ such that $\mu = 1+\x\nu$ and $\sigma=\x+\x^2\tau$ with $\nu,\tau\in\C[[\x]]$. The set of all pairs of such series $(\mu,\sigma)$ is naturally equipped with a semi-direct group structure called ``Riordan group" which can be univocally transported to the set of all Riordan matrices $M^{(\mu,\sigma)}$. The group multiplication is given by
$(\mu_1,\sigma_1)\rtimes(\mu_2,\sigma_2):=((\mu_1\circ \sigma_2)\mu_2,\sigma_1\times\sigma_2)$. The Riordan group also plays a rather important role in pure combinatorics. For instance it naturally appears in the umbral calculus setting \cite{Rom84} and is related in an obvious way to Sheffer sequences~\cite{HHS07,RKO73} since the exponential generating function of the ordinary generating function of each column satisfies the following condition~\cite{DPS$^+$04}: $\displaystyle \sum_{i\geq 0,j\geq 0}m_{i,j}\x^i \frac{\mathtt{y}^j}{j!}=\mu(\x)e^{\mathtt{y}\sigma(\x)}$. More recently the Riordan group also appeared in the new domain of combinatorial quantum physics, namely in the problem of normal ordering of boson strings \cite{CDP08,DC08,DPS$^+$04}. Let us say some words on the subject. A boson string is an element of the so-called Weyl algebra that is the quotient algebra $\C\{\mathtt{a},\mathtt{a}^{\dagger}\}/\langle \mathtt{a}\mathtt{a}^{\dagger}-\mathtt{a}^{\dagger}\mathtt{a}-1\rangle$ of the free algebra generated by two (distinct) letters $\mathtt{a}$ and $\mathtt{a}^{\dagger}$ by the two-sided ideal generated by noncommutative polynomials of the form $\mathtt{a}\mathtt{a}^{\dagger}-\mathtt{a}^{\dagger}\mathtt{a}-1$. Since the work of O. Ore \cite{Ore33}, it is well-known that $((\mathtt{a}^{\dagger})^i \mathtt{a}^j)_{i,j}$ is a Hamel basis for the Weyl algebra. Then a boson string is called to be in normal form if, and only if, it is written in this basis.  In papers~\cite{DC08,DPS$^+$04} the authors show that for an important class of boson strings $\Omega$, the coefficients $m_{i,j}$ of their decomposition in normal form $\Omega=\displaystyle\sum_{i,j}m_{i,j}(\mathtt{a}^{\dagger})^i \mathtt{a}^j$ define a Riordan matrix\footnote{In~\cite{DC08,DPS$^+$04} such matrices are called {\emph{matrix of unipotent substitution with prefunction operators}}.} $(m_{i,j})_{i,j}$. Using some properties of the Riordan group, the authors succeeded to compute, in an explicit way, the evolution operator $e^{\lambda \Omega}$, so important in quantum physics. In the paper~\cite{DPS$^+$04} was proved the following statement. 
\begin{center}
{\emph{Let $M$ be a Riordan matrix. Then for all $\lambda \in\C$, $M^{\lambda}$ also is a Riordan matrix.}}
\end{center}

In this paper, we develop a formal calculus on pairs of series $(\mup,\sigmap)$ such that $(1+\mup,\x+\sigmap)$ belongs to the Riordan group. More precisely it is shown that for every formal power series $f=\displaystyle\sum_{n\geq 0}f_n\x^n$ with coefficients in some field $\K$ of characteristic zero, $\displaystyle\sum_{n\geq 0}f_n (\mup,\sigmap)^{\rtimes n}$ defines an element of the {\emph{Riordan near algebra}} (which is nothing else than the Cartesian product of $\K[[\x]]$ with the maximal ideal generated by $\x$, and  equipped with some algebraic structure, see sect.~\ref{premieresection}), and where $(\mup,\sigmap)^{\rtimes n}$ is the usual $n$th power of $(\mup,\sigmap)$ with respect to the multiplicative law $\rtimes$ of the Riordan near algebra which extends the product of the Riordan group. In other terms, we extend and generalize the notion of formal substitution in $\K[[\x]]$ to the Riordan near algebra. This formal calculus plays a similar role to the usual holomorphic calculi for Banach or Fr\'echet algebras. In particular it makes possible to consider exponential, logarithm or inverse as series in monomials $(\mup,\sigmap)^{\rtimes n}$ in a way identical to those of $\K[[\x]]$. Using this formal calculus, we also prove \emph{en passant} the existence of another kind of generalized powers $(1+\mup,\x+\sigmap)^{\rtimes \lambda}$ using binomial series, where $(1+\mup,\x+\sigmap)$ belongs to the Riordan group and $\lambda \in\K$, such that $(1+\mup,\x+\sigmap)^{\rtimes\lambda}$ also is in the Riordan group and $(1+\mup,\x+\sigmap)^{\rtimes (\alpha+\beta)}=(1+\mup,\x+\sigmap)^{\rtimes \alpha}\rtimes(1+\mup,\x+\sigmap)^{\rtimes \beta}$.  This notion of generalized powers, although similar in appearance, is different from the one introduced for the Riordan matrices in~\cite{DPS$^+$04}. 
The matrix version in~\cite{DPS$^+$04} concerns the existence of generalized powers for elements of the Riordan group but seen as lower triangular infinite matrices, and therefore embedded in some algebra of infinite matrices. In these notes, we establish the same kind of statement but in another kind of algebras, namely, in a near algebra.

\section{The Riordan near algebra $\K[[\x]]\rtimes\mathfrak{M}$ of formal power series under multiplication and substitution}\label{premieresection}

\subsection{Basics on formal power series}\label{basics}
In this paragraph some basic and useful definitions and notations are provided. Many textbooks such as \cite{Bou70,Bou81,Stan97,Stan99} can be used as references on the subject.
The meaning of symbol ``$:=$'' is an equality by definition. The letter ``$\K$'' denote any field of characteristic zero and $\K[[\x]]$ is the $\K$-algebra of formal power series in one indeterminate $\x$. 
$\K[[\x]]$ is endowed with the usual $(\x)$-adic topology. In the sequel we suppose that each of its subsets is equipped with the induced topology. The $(\x)$-adic topology is equivalently given by the valuation $\nu$ the definition of which is recalled with some of its main properties. Let $+\infty\not\in\N$. Let $f=\displaystyle\sum_{n\geq 0}f_n\x^n$.
\begin{equation}
\nu(f):=\displaystyle\left \{
\begin{array}{ll}
+\infty & \mbox{if}\ f=0\ ,\\
\mathsf{inf}\{n\in \N : f_n\not=0\}& \mbox{otherwise}.
\end{array}\right .
\end{equation}
For all $f,g\in\K[[\x]]$,
\begin{enumerate}
\item $\nu(f+g)\geq \min\{\nu(f),\nu(g)\}$ with equality in case $\nu(f)\not=\nu(g)$;
\item $\nu(fg)=\nu(f)+\nu(g)$
\end{enumerate}
with the usual conventions $+\infty > n$ and $+\infty + n =n+\infty=+\infty$ for every $n\in\N$, $+\infty+\infty=+\infty$. In the sequel we also use the following conventions $(+\infty)n=n(+\infty)=+\infty$ for every $n\in\N\setminus\{0\}$ or $n=+\infty$ and $0n=n0=0$ for every $n\in \N$ or $n=+\infty$, $(+\infty)^n=+\infty$ for every $n\in\N\setminus\{0\}$. Sometimes we use the notation ``$n>0$'' that means ``$n\in\N\setminus\{0\}$ or $n=+\infty$'' when $n$ explicitly refers to the valuation of some series. \\
With the previous topology, $\K[[\x]]$ becomes a topological algebra (we put on $\K$ the discrete topology). In particular the multiplication is (jointly) continuous.\\
The coefficient $f_n$ of $\x^n$ in the series $f$ can also be denoted by $\langle f,\x^n\rangle$ so that $f$ should be written as the sum $\displaystyle\sum_{n\geq 0}\langle f,\x^n\rangle \x^n$. In particular, $\langle f,1\rangle$ is the constant term of the series $f$ which is also denoted $f(0)$. For every $n\in\N$ and $f\in\K[[\x]]$ we define as usually
\begin{equation}
f^n := \left \{
\begin{array}{lll}
1=x^0 & \mbox{if}&n=0\ ,\\
\underbrace{f\times \dots \times f}_{n\ \mathit{terms}}&\mbox{if}&n\geq 1\ .
\end{array}\right .
\end{equation}
(Here we adopt the symbol ``$\times$'' to emphasize the use of the multiplication in $\K[[\x]]$ but in what follows juxtaposition will be used.) Finally, when $R$ is a ring (with unit), $U(R)$ denotes its group of units: for instance, $U(\K[[\x]])$ is the set of series of order zero, \emph{i.e.}, the constant term is not null:  $U(\K[[\x]])=\{f\in\K[[\x]] : \langle f,1\rangle=f(0) \not=0\}$. We define the group of \emph{unipotent multiplications} (following the terminology of~\cite{CDP08,DC08,DPS$^+$04}) $\UP:=\{1+\x f : f\in \K[[\x]]\}$ which is a subgroup of $U(\K[[\x]])$.

\subsection{Ringoid of formal power series under substitution}

For a certain kind of formal power series, another product may be defined: the formal substitution. Roughly speaking if $\sigma$ is a series without constant term, that is $\sigma$ is an element of the ideal $(\x):=\x\K[[\x]]$, and $f=\displaystyle\sum_{n\geq 0}f_n\x^n$ is any series, then $f\circ \sigma:=\displaystyle\sum_{n\geq 0}f_n\sigma^n$ is a well-defined element of $\K[[\x]]$ called the \emph{substitution} of $f$ and $\sigma$. This operation is linear in its first variable but not in the second one. So under this substitution the ideal $(\x)$ does not behave as an algebra but as some more general structure called a ``ringoid''.

\subsubsection{Ringoids, composition rings, tri-operational algebras, and near algebras: a review}
\quad\\
In this short paragraph are recalled some basic definitions and facts about exotic algebraic structures equipped with two or three operations and closely related to the notion of substitution/composition, which is quite central in this work and therefore deserves a review. \\

A \emph{(right) near $\K$-algebra}~\cite{Bro68} over a field $\K$ is a $\K$-vector space $N$ equipped with an operation $\circ$ such that 
\begin{enumerate}
\item $(N,\circ)$ is a (non necessarily commutative) semigroup;
\item $(x+y)\circ z = (x\circ z)+(y\circ z)$;
\item $(\alpha x)\circ y=\alpha(x\circ y)$
\end{enumerate}
for every $x,y,z\in N$ and $\alpha\in\K$. In a right near algebra, the null vector $0$ of $N$ is a left zero for $\circ$, \emph{i.e.}, $0\circ x=0$, because $(N,+,0)$ is a group. Obviously every (associative) algebra (without a unit) can be seen as a right near algebra.  Let denote by $\mu : N\times N \rightarrow N$ the mapping $\mu(x,y):=x\circ y$. The semigroup  multiplication $\mu$ defines a right semigroup representation $\rho_{\mu}$ of $(N,\mu)$ on the vector space $N$,
\begin{equation}
\begin{array}{llll}
\rho_{\mu}: & N & \rightarrow & \End(N)\\
& y & \mapsto & \left (
\begin{array}{lll}
N & \rightarrow & N\\
x &\mapsto & \mu(x,y)
\end{array}\right )
\end{array}
\end{equation}
where $\End(V)$ is the $\K$-algebra of linear endomorphisms of the vector space $N$. In other terms, for every $x,y,z\in N$, $(\rho_{\mu}(x)\circ \rho_{\mu}(y))(z)=\rho_{\mu}(\mu(y,x))(z)$, and for every $\alpha,\beta\in \K$,
$\rho_{\mu}(x)(\alpha y + \beta z)=\alpha\rho_{\mu}(x)(y)+\beta\rho_{\mu}(x)(z)$. 
The notion of a two-sided ideal of a near algebra $N$ takes its immediate meaning in this setting: more precisely, a two-sided ideal $I$ of $N$ is a subvector space of $N$ such that for every $\mu(I\times R)\subset I \supset \mu(R\times I)$.  Moreover, when $(N,\circ)$ also has a (two-sided) unit $\mathtt{I}$, \emph{i.e.}, $(N,\circ, \mathtt{I})$ is a monoid, we also define the group of units of $N$, $U(N)$, as the group of invertible elements of the monoid $(N,\circ,\mathtt{I})$, that is,  $U(N):=\{x\in N: \exists y\in N,\ x\circ y=y\circ x=\mathtt{I}\}$. If $(N,\circ,\mathtt{I})$ is a monoid, then $\K\mathtt{I}$ does not lie necessarily in the center $Z(N):=\{x \in N: x\circ y = y\circ x\ ,\ \forall y\in N\}$, because in general it is not true that $(\alpha\mathtt{I})\circ x = x\circ(\alpha \mathtt{I})$. \\
As in algebras, an element $x$ of a near algebra $N$ is called a \emph{right zero divisor} if there is some non zero $y\in N$ such that $y\circ x=0$. Note that $0$ is not necessarily a right zero divisor. If $0$ also is a right zero for $\circ$, then a non zero $x \in N$ is called a \emph{left zero divisor} if there is some non zero $y\in N$, such that $x\circ y=0$. Again if $0$ is a two-sided zero for $\circ$, we say that $N$ is a \emph{domain} if there is no left or right zero divisor. \\ 
Suppose that $\K$ is a topological field, \emph{i.e.}, a field equipped with a topology such that $(\K,+,0)$ is a topological group and $(\K,\cdot,1)$ is a topological monoid, and that the near algebra $N$ is a topological $\K$-vector space for some given topology. We say that $N$ is a \emph{topological near algebra} if for every $y\in N$ the mappings $x \mapsto x\circ y$ and $x\mapsto y\circ x$ are continuous, that is, $\circ$ is separately continuous. \\ 

In the subsequent sections and subsections, we will consider near algebras in which $0$ also is a left zero for $\circ$, and therefore $(N,\circ, 0)$ is a semigroup with a (two-sided) zero.  Moreover the near algebras encountered will also have a two-sided neutral element $\mathtt{I}\not=0$ for $\circ$, in such a way that $(N,\circ,\mathtt{I},0)$ is a monoid with a zero.\\

The idea to consider generalized algebras endowed with three different operations, namely \emph{addition}, \emph{multiplication} and \emph{substitution}, can be traced back to the work of Menger~\cite{Men1,Men2,Men3} and Mannos~\cite{Man46} who considered the notion of tri-operational algebras. A \emph{tri-operational algebra} $R$ is a nonempty set together with three operations $+$, $.$ and $\circ$ - respectively called \emph{addition}, \emph{multiplication} and \emph{substitution} - and three mutually distinct distinguished elements $0$, $1$ and $\mathtt{I}$ that satisfy the following properties:
\begin{enumerate}
\item $(R,+,0,\cdot,1)$ is a commutative ring with unit $1$;
\item $(R,\circ,\mathtt{I})$ is a (non necessarily commutative) monoid with identity $\mathtt{I}$;
\item $(x+y)\circ z = (x\circ z) + (y\circ z)$;
\item $(x\cdot y)\circ z = (x\circ z)\cdot (y\circ z)$;
\item $1\circ 0=1$
\end{enumerate}
for every $x,y,z\in R$. A \emph{constant} of $R$ is an element $x \in R$ such that $x=x\circ 0$. In particular, $1$ and $0$ are both constants, the set $CR$ of all constants is a commutative ring, called the \emph{ring of constants of $R$}, and, it can be easily checked that $(R,+,0,.,1)$ is a $CR$-algebra with unit $1$ (in particular $(R,+,0)$ is a unitary $CR$-module). For instance, if $A$ is a commutative ring with unit, then $A[\mathtt{x}]$ is a tri-operational algebra under the usual operations with $\mathtt{I}=\mathtt{x}$. Conversely, if $R$ is a tri-operational algebra, then the set $\Pi(R)$of all elements of $R$ of the form $\alpha_0+\alpha_1\cdot\mathtt{I}+\alpha_2\cdot\mathtt{I}+\cdots+\alpha_n\cdot\mathtt{I}^n$, for $n\in \N$, where, for every $i$, $\alpha_i\in CR$ and $\mathtt{I}^i:=\underbrace{\mathtt{I}\cdot \mathtt{I}\cdots\mathtt{I}}_{i\ \mathit{factors}}$ if $i\not=0$, is a tri-operational algebra for the operations induced by $R$ on $CR$, and therefore a \emph{tri-operational subring} (in an obvious way) of $R$, called the \emph{algebra of polynomials of $R$}. Note that $\Pi(R)$ is not necessarily isomorphic to $R[\mathtt{x}]$ because $\mathtt{I}$ may not be algebraically free over the ring $CR$. \\

In~\cite{Adl62}, Adler generalized the concept of tri-operational algebra. A \emph{composition ring} $R$ is a ring $(R,+,0,\cdot)$ (possibly without a unit) equipped with an operation $\circ$ such that 
\begin{enumerate}
\item $(R,\circ)$ is a (non necessarily commutative) semigroup;
\item $(x+y)\circ z=(x\circ y)+(x\circ z)$;
\item $(x\cdot y)\circ z=(x\circ z)\cdot(x\circ y)$.
\end{enumerate}
A tri-operational algebra is nothing else than a composition ring with a multiplicative unit $1\not=0$ and with a (two-sided) unit for composition $\mathtt{I}$, $\mathtt{I}\not=0$, $\mathtt{I}\not=1$, such that $1\circ 0=1$. The last fact is possible only if there is at least one element of $R$ that is not a zero-divisor of the carrier ring $(R,+,0,\cdot,1)$. A \emph{constant} of $R$ is an element $x \in R$ such that $x \circ y = x$ for every $y\in R$. The set $\mathtt{Found}\ R$ of all constants of $R$ is called the \emph{foundation} of $R$, and it is a composition subring of $R$. In particular, $(\mathtt{Found}\ R,+,0,\cdot)$ is a commutative sub-ring (possibly without a unit) of $(R,+,0,\cdot)$. \\

Finally, Iskander~\cite{Isk71} introduced the following kind of structures. A \emph{ringoid} $(R,+,\cdot,\circ)$ is nonvoid set with three operations $+$, $\cdot$ and $\circ$ such that 
\begin{enumerate}
\item $(R,+)$ is a commutative semigroup;
\item $(R,\cdot)$ is a commutative semigroup;
\item $(x+y)\cdot z=(x\cdot z)+(x\cdot z)$;
\item $(R,\circ)$ is a (non necessarily commutative) semigroup;
\item $(x+y)\circ z=(x\circ z)+(y\circ z)$;
\item $(x\cdot y)\circ z=(x\circ z)\cdot(y\circ z)$
\end{enumerate}
for every $x,y,z\in R$. The first three axioms mean that $(R,+,\cdot)$ is a commutative semiring (without $0$). A \emph{ringoid with units $0$, $1$ and $\mathtt{I}$} is a ringoid with three mutually distinct elements $0$, $1$ and $\mathtt{I}$ that are (two-sided) neutral elements for respectively $+$, $\cdot$ and $\circ$, such that $0\circ x=0$ and $1\circ x=1$ for every $x\in R$. Note that $(R,+,0,\cdot,1)$ becomes a semiring (without the usual requirement that $x\cdot 0=0$). A composition ring is a ringoid with an element $0\in R$ such that $(R,+,0)$ becomes a (commutative) group. In this case, $(R,+,0,\cdot)$ is a ring and $x\cdot 0=0$ for every $x \in R$ (in other terms, $(R,\cdot,0)$ is a semigroup with a zero).\\
A \emph{topological ringoid} is a ringoid $R$ together with a topology such that $(R,+,\cdot)$ is a topological semiring and $\circ$ is separately continuous.\\

In what follows we will present a ringoid  $(R,+,0,\cdot,\circ,\mathtt{I})$ such that $R$ is a composition ring ($0$ is the identity of the group $(R,+,0)$) with a two-sided identity $\mathtt{I}$ for the operation of substitution $\circ$ such that 
$(R,\circ,\mathtt{I},0)$ is a (non necessarily commutative) monoid with a two-sided\footnote{The fact that $0$ is a left zero for $\circ$ is true in any composition ring since $(R,+,0)$ is a group.} zero $0$, \emph{i.e.}, $x\circ 0 = 0\circ x=0$ for every $x \in R$. Note that in this case, if the ring $(R,+,0,\cdot)$ has no multiplicative unit (which will be the case), then the foundation of the composition ring is reduced to $(0)$ since $x$ is a constant if, and only if, $x \circ y=x$ for every $y\in R$, so, in particular, $x\circ 0=x$ but, because we assume that $0$ is a right-zero for the composition, $x=0$.

\subsubsection{Ringoid of formal power series under substitution}
\quad\\
Let $\gotM := (\x)=\x\K[[\x]]$ be the principal ideal generated by $\x$. It is the unique maximal ideal of $\K[[\x]]$ and it also generates the $(\x)$-adic topology. Due to the definition of $\gotM$ any of its elements has a positive valuation (since the constant term is equal to zero). The operation ``$\circ$'' of formal substitution of power series turns $\gotM$ into a (noncommutative) monoid whose (two-sided) identity is $\x$. If $\sigma\in \gotM$ and $n\in\N$, we may define
\begin{equation}
\sigma^{{\circ n}}:=\left \{
\begin{array}{lll}
\x & \mbox{if} &n=0\ ,\\
\underbrace{\sigma \circ \dots \circ \sigma}_{n\ \mathit{terms}} & \mbox{if}&n\geq 1\ .
\end{array}\right .
\end{equation}
The operation of right substitution by an element $\sigma\in\gotM$ on $\K[[\x]]$ defines a $\K$-algebra endomorphism, that is,
\begin{equation}
\begin{array}{lll}
\K[[\x]] & \rightarrow & \K[[\x]]\\
f&\mapsto & f\circ \sigma
\end{array}
\end{equation}
is a $\K$-algebra endomorphism of $\K[[\x]]$. Such an endomorphism is an automorphism if, and only if, $\nu(\sigma)=1$ (or, equivalently, the coefficient $\langle \sigma,\x\rangle$ of $\x$ in $\sigma$ is non zero). More generally we can prove that in many cases the above endomorphism is one-to-one.
\begin{lemma}
Let $\sigma \in \gotM\setminus\{0\}$. Then, right substitution by $\sigma$ is one-to-one.
\end{lemma}
\begin{proof}
Suppose the contrary and let $f=\displaystyle\sum_{n\geq 0}f_n\x^n\in\K[[\x]]\setminus\{0\}$ such that $f\circ \sigma=0$. Let $m:=\nu(f)\not=\infty$ and $\ell:=\nu(\sigma)>0$. By assumption, we have $m\geq 1$ and $f_m\sigma_{\ell}^m=0$ which contradicts the fact that $\K$ is a field.
\end{proof}

This lemma implies that for every $\sigma,\tau\in\gotM$, if $\sigma\circ \tau=0$ then $\sigma=0$ or $\tau=0$. Indeed if $\tau\not=0$, then by the previous lemma, $\sigma=0$. If $\sigma\not=0$, then $\tau=0$ and we are done ($\sigma\circ 0=\sigma(0)=0$ because $\nu(\sigma)>0$) or $\tau\not=0$, but the later case contradicts the lemma.\\

The group of invertible elements of the monoid $\gotM$ is then precisely given by $\{\sigma \in \gotM : \langle \sigma,\x\rangle\not=0\}$, that is, the set of series that ``begin exactly by some (nonzero) multiple of $\x$''. With the usual addition of formal power series, $\gotM$ becomes a right near algebra (without zero divisor), with $\mathtt{x}$ has the two-sided identity for $\circ$, which is not an algebra. Indeed, for instance,  $\x^2 \circ (\x-\x)=0$ but $\x^2 \circ \x +\x^2\circ(-\x)=2\x^2\not=0$ (since $\K$ is field of characteristic zero), or also $\x^2\circ (2\x)=4\x^2\not=(2\x^2)\circ\x=2\x^2$. The group of units $U(\gotM)$ of the algebra $\gotM$ is the group of invertible elements of the corresponding monoid, and, $\US:=\{\x+\x^2 f : f\in\K[[\x]]\}$ is a subgroup of  $U(\gotM)$, called the group of \emph{unipotent substitutions}. Note also that $0$ is a two-sided zero for the operation of substitution. \\
Because right composition is an algebra endomorphism, it can be easily checked that $(\gotM,+,0,\times,\circ,\mathtt{x})$ also is a ringoid, with $0$ neutral for $+$, and a two-sided zero for $\circ$. 
\begin{rem}
This structure can be extended to the whole $\K[[\mathtt{x}]]$ as follows. Let $\omegaup\not\in \K[[\mathtt{x}]]$. We extend addition and multiplication to $\K[[\x]]\cup\{\omegaup\}$ by 
\begin{equation}
\begin{array}{lllll}
\omegaup+f&=&f+\omegaup&=&\omegaup\ ,\\
\omegaup f&=&f\omegaup&=&\omegaup
\end{array}
\end{equation}
for every $f \in \K[[\x]]\cup\{\omegaup\}$. Then,  $(\K[[\x]]\cup\{\omegaup\},+,0,\cdot,1)$ is a commutative semiring (with a zero $\omegaup$ for both addition and multiplication). We also extend $\circ$ to $\K[[\x]]$ by $\omegaup\circ f = \omegaup$ for every $f \in \K[[\x]]\cup\{\omegaup\}$, and, for every $f \in \K[[\x]]$ and $g\in \K[[\x]]\cup\{\omegaup\}$, 
\begin{equation}
f\circ g = \left \{
\begin{array}{lll}
f\circ g &\mbox{if} & g\in\gotM\ ,\\
\omegaup & \mbox{if}& g\not\in\gotM\ .
\end{array}
 \right .
\end{equation}
In particular, $f\circ\omegaup=\omegaup$ for every $f \in \K[[\x]]$, because $\omegaup\not\in\gotM$. Then, $(\K[[\x]]\cup\{\omegaup\},\circ,\x)$ is a monoid with a zero $\omegaup$, and $(\K[[\x]]\cup\{\omegaup\},+,0,\cdot,1,\circ,\x)$ is a ringoid. 
\end{rem}

When we put on $\K$ the discrete topology and on $\gotM$ the subspace topology, the later is immediately seen as a Hausdorff (since metrizable) topological vector space on the former.
\begin{lemma}\label{continuityofcomposition}
The formal substitution is separately continuous. More precisely, for every $\sigma_r\in \gotM$,
\begin{equation}
\begin{array}{llll}
s^{(r)}_{\sigma_r}: & \gotM & \rightarrow & \gotM\\
&\sigma & \mapsto & \sigma\circ\sigma_r
\end{array}
\end{equation}
is a continuous linear endomorphism and for every $\sigma_l \in \gotM$,
\begin{equation}
\begin{array}{llll}
s^{(l)}_{\sigma_l}: & \gotM & \rightarrow & \gotM\\
&\sigma & \mapsto & \sigma_l\circ\sigma
\end{array}
\end{equation}
is a continuous (nonlinear) mapping.
\end{lemma}
\begin{proof}
Left to the reader.
\end{proof}
It follows that $\gotM$ is both a topological near algebra and a topological ringoid.

\subsection{Near algebra $\K[[\x]]\rtimes\mathfrak{M}$}
On the set-theoretic cartesian product $\K[[\x]]\times\gotM$ it is possible to define a natural structure of right near  $\K$-algebra. This near algebra is denoted by $\K[[\x]]\rtimes\gotM$ and called the \emph{Riordan near algebra}. The additive structure of the underlying $\K$-vector space is the usual one given by the direct sum. The multiplication is defined by the following rule for each $(\mu_1,\sigma_1),(\mu_2,\sigma_2)\in \K[[\x]]\times \gotM$
\begin{equation}\label{def_prod_semi_direct}
(\mu_1,\sigma_1)\rtimes(\mu_2,\sigma_2):=((\mu_1 \circ \sigma_2)\mu_2,\sigma_1\circ\sigma_2)\ .
\end{equation}
It is left to the reader to check that this formula defines a noncommutative monoid multiplication (and in particular an associative binary law) with $(1,\x)$ as its identity element. Likewise in $\gotM$, $(0,0)$ is a right (and therefore a two-sided) zero for $\rtimes$. As easily one can prove that the group of units of $\K[[\x]]\rtimes\gotM$ is the semi-direct product of the group of units of each (near) algebra. More precisely, $U(\K[[\x]]\rtimes\gotM)=U(\K[[\x]])\rtimes U(\gotM)$ where $\rtimes$ is defined as in the formula~(\ref{def_prod_semi_direct}). Moreover the semi-direct group $\UP\rtimes\US$ is a subgroup of $U(\K[[\x]]\rtimes\gotM)$. It is called the \emph{Riordan group} as originally introduced and studied in~\cite{SGW$^+$91}. The near algebra $\K[[\x]]\rtimes\gotM$ is far from being a domain because for instance every nonzero element of the two-sided ideal $(0)\times\gotM$ is a right zero divisor and every nonzero element of the right ideal $\K[[\x]]\rtimes(0)$ is a left zero divisor:
$(\mu,0)\rtimes(0,\sigma)=(0,0)$.  \\
The near algebra $\gotM$ may be identified with a two-sided ideal of
$\K[[\x]]\rtimes\gotM$ by the natural injection
\begin{equation}
\begin{array}{lll}
\gotM&\rightarrow & \K[[\x]]\rtimes\gotM\\
\sigma & \mapsto & (0,\sigma)
\end{array}
\end{equation}
whereas $\K[[\x]]$ can only be identified separately as a submonoid  and as a subvector space of $\K[[\x]]\rtimes \gotM$ by the respective one-to-one homomorphisms (the first one is a morphism of monoids, and the second one is a linear mapping)
\begin{equation}
\begin{array}{lll}
\K[[\x]] & \rightarrow & \K[[\x]]\rtimes\gotM\\
\mu  &\mapsto & (\mu,\x)
\end{array}
\end{equation}
and
\begin{equation}
\begin{array}{lll}
\K[[\x]] & \rightarrow & \K[[\x]]\rtimes\gotM\\
\mu  &\mapsto & (\mu,0)
\end{array}
\end{equation}
It obviously holds that each of these embeddings is also continuous ($\K[[\x]]\rtimes\gotM$ has the product topology) and both vector spaces $(0)\times\gotM$ and $\K[[\x]]\times (0)$ are closed in the Riordan near algebra.\\

We define the generalized product in a usual fashion. For each $n\in\N$ and $(\mu,\sigma)\in\K[[\x]]\rtimes\gotM$, we put
\begin{equation}
(\mu,\sigma)^{\rtimes n}:=\displaystyle\left \{
\begin{array}{lll}
(1,\x) & \mbox{if}& n=0\ ,\\
\underbrace{(\mu,\sigma)\rtimes\dots\rtimes(\mu,\sigma)}_{n\ \mathit{terms}} & \mbox{if} & n\geq 1\ .
\end{array}
\right .
\end{equation}
The following easy lemma will be useful in the sequel.
\begin{lemma}\label{calculexplicitedelasommegeneraliseeduproduitsemidirect}
For each $(\mu,\sigma)\in\K[[\x]]\rtimes\gotMp$ and $n\in\N$,
\begin{equation}
(\mu,\sigma)^{\rtimes n}=\left \{
\begin{array}{lll}
(1,\x) & \mbox{if} & n=0\ ,\\
\displaystyle \left ( \prod_{k=1}^n (\mu\circ\sigma^{\circ(k-1)}),\sigma^{{\circ n}}\right )&\mbox{if} & n\geq 1\ .
\end{array}
\right .
\end{equation}
In particular if $\sigma=0$, then
\begin{equation}
(\mu,0)^{\rtimes n}=\left \{
\begin{array}{lll}
(1,\x) & \mbox{if} & n=0\ ,\\
(\mu\mu(0)^{n-1},0)&\mbox{if} & n\geq 1\ .
\end{array}
\right .
\end{equation}
(Under the convention $\alpha^0:=1$ for every $\alpha\in\K$ in such a way that $(\mu,0)^{\rtimes 1}=(\mu\mu(0)^{0},0)=(\mu,0)$ even for $\mu(0)=0$.)\\
If $\mu=0$, then
\begin{equation}
(0,\sigma)^{\rtimes n}=\left \{
\begin{array}{lll}
(1,\x) & \mbox{if} & n=0\ ,\\
(0,\sigma^{{\circ n}})&\mbox{if} & n\geq 1\ .
\end{array}
\right .
\end{equation}
\end{lemma}
\begin{proof} 
Omitted.
\end{proof}

\subsubsection{Topological considerations}
\quad\\
In the remainder of the paper, we suppose that the underlying set $\K[[\x]]\times\gotM$ of the near algebra $\K[[\x]]\rtimes \gotM$ is equipped with the product topology in such a way that the underlying vector space is a Hausdorff (since the topology is metrizable) topological vector space (when is put on $\K$ the discrete topology). Regarding the multiplicative structure, the following result is proved.
\begin{lemma}\label{continuitesepareedelamultiplication}
$\rtimes$ is separately continuous. More precisely, for every $(\mu_r,\sigma_r)\in\K[[\x]]\rtimes\gotM$, the mapping
\begin{equation}
\begin{array}{llll}
R_{(\mu_r,\sigma_r)}: & \K[[\x]]\rtimes \gotM&\rightarrow &\K[[\x]]\rtimes\gotM\\
& (\mu,\sigma)&\mapsto & (\mu,\sigma)\rtimes(\mu_r,\sigma_r)
\end{array}
\end{equation}
is a continuous linear endomorphism, and for every $(\mu_l,\sigma_l)\in\K[[\x]]\rtimes\gotM$,
\begin{equation}
\begin{array}{llll}
L_{(\mu_l,\sigma_l)}:& \K[[\x]]\rtimes \gotM&\rightarrow &\K[[\x]]\rtimes\gotM\\\
& (\mu,\sigma)&\mapsto & (\mu_l,\sigma_l)\rtimes(\mu,\sigma)
\end{array}
\end{equation}
is a (nonlinear) continuous mapping.
\end{lemma}
Before we achieve the proof of this result, we need another easy lemma.
\begin{lemma}\label{ordredunecompositon}
Let $(\mu,\sigma)\in \K[[\x]]\times\gotM$. Then $\nu(\mu\circ\sigma)=\nu(\mu)\nu(\sigma)$. In particular, for every $n\in\N$, $\nu(\sigma^{{\circ n}})=\nu(\sigma)^n$. (Under the conventions recalled in subsect.~\ref{basics}: $(+\infty)n=n(+\infty)=+\infty$ if $n\in \N\setminus\{0\}$ or $n=+\infty$, $0n=n0=0$ if $n\in\N$ or $n=+\infty$, $(+\infty)^n=+\infty$ for every $n\in\N\setminus\{0\}$.)
\end{lemma}
\begin{proof}
Let us begin to prove that $\nu(\mu\circ\sigma)=\nu(\mu)\nu(\sigma)$.
\begin{enumerate}
\item Suppose that $\mu=0$. Then $\mu\circ \sigma=0$. Since $\nu(0)=+\infty$, $\nu(\sigma)>0$ and $(+\infty)n=n(+\infty)=+\infty$ for every $n>0$, the result follows;
\item Suppose that $\mu\not=0$. If $\sigma=0$, then $\nu\circ \sigma=\mu(0)$. Now if $\mu(0)=0$, that is, $\nu(\mu)>0$, then $\nu(\mu(0))=+\infty=\nu(\mu)(+\infty)=\nu(\mu)\nu(0)$. If $\mu(0)\not=0$, that is, $\nu(\mu)=0$, then $\nu(\mu(0))=0=\nu(\mu)0=\nu(\mu)\nu(\sigma)$. Finally let suppose that $\sigma\not=0$. Because $\mu\not=0$, there is a $n_0\in\N$ such that $n_0=\nu(\mu)$ and $\mu=\displaystyle\sum_{n\geq n_0}\mu_n \x^n$ with $\mu_{n_0}\not=0$. By definition, $\mu\circ\sigma=\displaystyle\sum_{n\geq n_0}\mu_n\sigma^n$. But $\nu(\sigma^n)=n\nu(\sigma)$ for every $n\in \N$. Since $\nu(\sigma)>0$, for all $m>n$, $\nu(\sigma^m)>\nu(\sigma^n)$ and in particular for every $n>n_0=\nu(\mu)$, $\nu(\sigma^n)>\nu(\sigma^{n_0})=n_0\nu(\sigma)=\nu(\mu)\nu(\sigma)$ and for every $n<n_0$, $\mu_n\sigma^n=0$.
\end{enumerate}
Now let us prove the second statement of the lemma. Let $\sigma\in \gotM$ and $n\in\N$.
\begin{enumerate}
\item Suppose that $\sigma=0$. Therefore $0^{{\circ n}}=\displaystyle\left \{
\begin{array}{lll}
\x & \mbox{if} & n=0\ ,\\
0 & \mbox{if} & n\in\N\setminus\{0\}
\end{array}\right .$ which implies that $\nu(0^{{\circ n}})=\displaystyle\left \{
\begin{array}{lll}
1 & \mbox{if} & n=0\ ,\\
+\infty & \mbox{if} & n\in\N\setminus\{0\}
\end{array}
\right .$. The expected result follows;
\item Suppose that $\sigma\not=0$ (that is to says that $\nu(\sigma)\in\N\setminus\{0\}$). $\nu(\sigma^{\circ 0})=\nu(\x)=1=\nu(\sigma)^0$. Suppose by induction that $\nu(\sigma^{{\circ n}})=\nu(\sigma)^n$. Then $\nu(\sigma^{\circ (n+1)})=\nu(\sigma^{{\circ n}}\circ\sigma)=\nu(\sigma^{{\circ n}})\nu(\sigma)$ (according to the first statement of the lemma) $=\nu(\sigma)^{n+1}$ by induction.
\end{enumerate}
\end{proof}

\begin{proof} (of lemma~\ref{continuitesepareedelamultiplication})
\begin{enumerate}
\item Let us begin with $R_{(\mu_r,\sigma_r)}$: it is already known to be linear. Therefore we only need to check continuity at zero. Let $((\mu_n,\sigma_n))_{n\in\N}$ be a sequence of elements of $\K[[\x]]\rtimes\gotM$ converging to $(0,0)$, which, by definition of the product topology, is equivalent to $\nu(\mu_n)$ and $\nu(\sigma_n)$ both converge to $+\infty$. But $(\mu_n,\sigma_n)\rtimes (\mu_r,\sigma_r)=((\mu_n \circ \sigma_r)\mu_r,\sigma_n\circ\sigma_r)$. Now $\nu((\mu_n\circ \sigma_r)\mu_r)=\nu(\mu_n\circ\sigma_r)+\nu(\mu_r)=\nu(\mu_n)\nu(\sigma_r)+\nu(\mu_r)$, according to lemma~\ref{ordredunecompositon}. Because $\nu(\sigma_r)>0$, it follows that $\nu((\mu_n,\sigma_n)\rtimes (\mu_r,\sigma_r))$ converges to $+\infty$ as $n\rightarrow+\infty$. So the first component of $R_{(\mu_r,\sigma_r)}(\mu_n,\sigma_n)$ converges to zero as $n\rightarrow+\infty$. Moreover $\nu(\sigma_n\circ\sigma_r)=\nu(\sigma_n)\nu(\sigma_r)$, and for the same reason as the first component, the second component also converges to zero. By definition of the product topology of two metrizable topologies, the result is proved;
\item Let us explore the case of $L_{(\mu_l,\sigma_l)}$: we begin to prove that the following mapping is continuous.
\begin{equation}
\begin{array}{llll}
\ell_{\mu_l}: & \gotM & \rightarrow & \K[[\x]]\\
& \sigma & \mapsto & \mu_l \circ \sigma
\end{array}
\end{equation}
Let $(\sigma_n)_{n\in\N}\in\gotM^{\N}$ which converges to $\sigma\in \gotM$. We should prove that $\mu_l \circ \sigma_n \rightarrow \mu_l\circ\sigma$ which actually is immediate. Therefore $\ell_{\mu_l}$ is continuous. Now we need a general result recalled below.\\
Let $X_1,X_2,Y_1,Y_2$ and $Z$ be topological spaces and $h:Y_1\times Y_2\rightarrow Z$ be a continuous mapping ($Y_1\times Y_2$ with the product topology). Let $f_i : X_i \rightarrow Y_i$ for $i=1,2$ be continuous mappings. Then the mapping
\begin{equation}
\begin{array}{llll}
f_1 \otimes_h f_2: & X_1 \times X_2 &\rightarrow&Z\\
& (x_1,x_2) & \mapsto & h(f_1(x_1),f_2(x_2))
\end{array}
\end{equation}
is also continuous ($X_1\times X_2$ with the product topology). \\
It is possible to take advantage of this later general statement in our case because the first coordinate function of $L_{(\mu_l,\sigma_l)}$ is equal to $\Id\otimes_{\times}\ell_{\mu_l}$ (where $\Id$ stands for the identity mapping of $\K[[\x]]$ and $\times$ for the usual multiplication of $\K[[\x]]$ which is known to be continuous). Finally the second coordinate function of $L_{(\mu_l,\sigma_l)}$ is
\begin{equation}
\begin{array}{lll}
 \K[[\x]]\rtimes\gotM & \rightarrow & \gotM\\
 (\mu,\sigma) &\mapsto & \sigma_l \circ \sigma
\end{array}
\end{equation}
which is continuous by lemma~\ref{continuityofcomposition} since for every $(\mu,\sigma)$ it is equal to $s_{\sigma_l}^{(l)}(\sigma)$. By definition of the product topology, continuity of both coordinate functions implies the continuity of $L_{(\mu_l,\sigma_l)}$ itself.
\end{enumerate}
\end{proof}

In what follows we consider convergent series of elements of $\K[[\x]]$. In some cases, convergence actually implies summability, that is to say that the sum of the series does not depend on the order of summation. A formal power series $f\in\K[[\x]]$ is said {\emph{topologically nilpotent}} if, and only if, $\displaystyle\lim_{n\rightarrow+\infty}f^n=0$. For such series the following assertion holds.
\begin{lemma}\label{topological_nilpotence}
Let $f\in\K[[\x]]$ be a topologically nilpotent series. Then for every sequence of scalars $(\alpha_n)_{n\in\N}$, the family $(\alpha_n f^n)_{n\in\N}$ is summable.
\end{lemma}
\begin{proof}
According to theorem~10.4 \cite{War93} since $\K[[\x]]$ is a complete Hausdorff commutative group, it is sufficient to prove that $(\alpha_n f^n)_{n\in\N}$ satisfies Cauchy's condition, namely for every neighborhood $V$ of zero in $\K[[\x]]$ there exists a finite subset $J_{V}$ of $\N$ such that for every finite subset $K$ of $\N$ disjoint from $J_V$, $\displaystyle\sum_{n\in K}\alpha_n f^n \in V$. So let $I$ be a finite subset of $\N$. Let $V_I:=\{g\in\K[[\x]] : \langle g, \x^n\rangle = 0\ \mbox{for every}\ n\in I\}$ be a neighborhood of zero. Because $f$ is topologically nilpotent, for every $m\in\N$ there exists $n_m\in \N$ such that for every $n>n_m$, $\nu(f^n)>m$. Therefore in particular for every $k\leq m$ and every $n>n_m$,
$\langle \alpha_n f^n,\x^k\rangle=0$. Thus for every finite subset $K$ of $\N$ disjoint from $\{0,\dots,n_{\max{\{I\}}}\}$, $\langle \alpha_k f^k,\x^i\rangle=0$ for every $k\in K$ and $i\in I$. Then $\displaystyle\sum_{k\in K}\alpha_k f^k\in V_I$.
\end{proof}

More generally the same lemma holds for $\K[[\x]]\rtimes\gotM$ in place of $\K[[\x]]$, where we call $(\mu,\sigma)$ topologically nilpotent if, and only if, $\displaystyle\lim_{n\rightarrow\infty}(\mu,\sigma)^{\rtimes n}=0$.
\begin{lemma}\label{topological_nilpotence_in_dibruno}
Let $(\mu,\sigma)\in \K[[\x]]\rtimes\gotM$ be topologically nilpotent. Then for every sequence of scalars $(\alpha_n)_n$, the family $(\alpha_n(\mu,\sigma)^{\rtimes n})_{n\in\N}$ is summable.
\end{lemma}

\begin{proof}
Because $\K[[\x]]\rtimes\gotM$ is a complete Hausdorff commutative group for the product topology, it is sufficient to prove that the summability holds componentwise. It is obvious to prove that $\sigma$ is topologically nilpotent in $\gotM$ that is to say that $\displaystyle\lim_{n\rightarrow\infty}\sigma^{{\circ n}}=0$. Therefore by a trivial variation of lemma~\ref{topological_nilpotence} it implies that $(\alpha_n \sigma^{{\circ n}})_n$ is summable in $\gotM$ (note that $\gotM$ is easily seen to be closed in $\K[[\x]]$ and therefore is complete). It remains to prove that the family $(\displaystyle u_n)_n$ is summable in $\K[[\x]]$ where
$u_n=\displaystyle\left \{ \begin{array}{lll}
\alpha_0 & \mbox{if} & n=0\ ,\\
\alpha_n\displaystyle\prod_{k=1}^{n}(\mu\circ\sigma^{\circ(k-1)})&\mbox{if} & n\geq 1
\end{array}\right .$. Because $(\mu,\sigma)$ is topologically nilpotent, it implies that $\displaystyle\lim_{n\rightarrow\infty}u_n=0$. So for every $m\in \N$ there exists $n_m\in\N$ such that for every $n>n_m$, $\langle u_n,\x^k\rangle=0$ for all $k\leq m$. The conclusion follows by a slight adaptation of the proof of lemma~\ref{topological_nilpotence}.
\end{proof}
In the previous proof we saw that $\gotM$ is a complete (as a vector space). We say that $\sigma\in\gotM$ is \emph{topological nilpotent} if, and only if, $\displaystyle\lim_{n\rightarrow\infty}\sigma^{\circ n}=0$. Then, by a minor modification of the proof of lemma~\ref{topological_nilpotence}, we easily deduce the following.
\begin{lemma}\label{topological-nilpoence-in-gotM}
Let $\sigma\in \gotM$ be topologically nilpotent. Then for every sequence of scalars $(\alpha_n)_n$, the family $(\alpha_n\sigma^{\circ n})_{n\in\N}$ is summable.
\end{lemma} 
Note that by the identification of $\gotM$ with $(0)\times\gotM$, $\sigma$ is topologically nilpotent in $\gotM$ if, and only if, $(0,\sigma)$ is topologically nilpotent in $\K[[\x]]\rtimes\gotM$.

\section{Formal calculus on the Riordan near algebra}\label{formal-power-series-calculus}

\subsection{Introduction}

The goal of this section is to develop a formal calculus on the Riordan near algebra. The idea is to extend the notion of formal substitution to this new algebraic framework: given a series $f=\displaystyle\sum_{n\geq 0}f_n\x^n$ and some particular element $(\mup,\sigmap)$ of the Riordan near algebra, it will appear that the series $\displaystyle\sum_{n\geq 0}f_n (\mup,\sigmap)^{\rtimes n}$ obtained by substitution of $\x$ by $(\mup,\sigmap)$ is convergent in the Riordan near algebra just as $\displaystyle \sum_{n\geq 0}f_n \sigma^n$ defines a formal power series whenever $\sigma\in\gotM$. The two-sided ideal $\Kxp\rtimes\gotMp$ of the Riordan near algebra, given by pairs of series $(\mup,\sigmap)$ of orders respectively positive and strictly greater than one,  plays the same role as the ideal $\gotM$ for the usual substitution. This formal calculus allows us to define exponential, logarithm and inverse series in the Riordan near algebra by using their usual formal power series versions where monomials in $\x$ are replaced by powers of $(\mup,\sigmap)$. Nevertheless, due to the lack of commutativity and left-distributivity of the Riordan near algebra, the usual properties of these series fail to be true in the new setting. For instance the inverse series $\displaystyle\sum_{n\geq 0}(\mup,\sigmap)^{\rtimes n}$ is not the inverse of $(1,\x)-(\mup,\sigmap)$ in the Riordan near algebra. It will be the main objective of sect.~\ref{sect4} to provide a convenient algebraic setting in which these series play their expected roles.

\subsection{Power series of elements of $\Kxp\rtimes\gotMp$}\label{calcul_formel}

Generally speaking a formal power series $f:=\displaystyle \sum_{n\geq 0}f_n\x^n$ is said to {\emph{operate}} on an element $a$ of a topological (associative) algebra $A$ (with unit $1_A$) if and only if the series $\displaystyle\sum_{n\geq 0}f_n a^n$ (with $a^0:=1_A$ and  $a^{n+1}:=aa^{n}$) converges in the topology of $A$. If each element of a given subset $S\subseteq  \K[[\x]]$ operates on $a$, we say that $S$ {\emph{operates on $a$}}. Finally if $S$ operates on each element of $T\subseteq A$, then we say that {\emph{$S$ operates on $T$}}. In this section we prove that there exists a two-sided ideal of $\K[[\x]]\rtimes\gotM$ on which every element of $\K[[\x]]$ operate. This allows us to define a formal calculus on the Riordan near algebra.\\

We define $\gotMp:=\{\sigma\in\gotM : \nu(\sigma)>1\}$, or in other terms, an arbitrary element $\sigma$ of $\gotMp$ takes the form $\sigma = \alpha \x^2+\x^3 f$, where $f\in\K[[\x]]$. The set $\gotMp$ is a two-sided ideal of the near algebra $\gotM$. Indeed, $\nu(\sigma+\tau)\geq \min\{\nu(\sigma),\nu(\tau)\}>1$ and $\alpha \sigma\in \gotMp$ for every $\sigma,\tau\in\gotMp$ and every $\alpha\in \K$. Now let $\sigma\in\gotM$ and $\sigmap\in\gotMp$, then $\nu(\sigma\circ\sigmap)=\nu(\sigma)\nu(\sigmap)>1$ and $\nu(\sigmap\circ\sigma)=\nu(\sigmap)\nu(\sigma)>1$ (since $\nu(\sigmap)\geq 1$) which ensure that $\gotMp$ is a two-sided ideal of $\gotM$.\\

In a similar way we define $\Kxp:=\gotM$. We use another name for $\gotM$ because in the subsequent part of this paper its multiplicative structure will be important, at least more important than its compositional structure. $\Kxp$ is a two-sided ideal of
$\K[[\x]]$. Indeed, $\nu(\lambda + \mu)\geq \min\{\nu(\lambda),\nu(\mu)\}> 0$ and $\alpha \mu \in \Kxp$ for every $\lambda,\mu\in\Kxp$ and $\alpha\in\K$. Now let $\mu\in\K[[\x]]$ and $\mup\in\Kxp$. Then $\nu(\mu\mup)=\nu(\mu)+\nu(\mup)>0$ which ensures that $\Kxp$ is an ideal of the commutative algebra $\K[[\x]]$.\\

Now let show that $\Kxp\rtimes\gotMp$ is itself a two-sided ideal of $\K[[\x]]\rtimes\gotM$. Obviously regarding the vector space structure, there is nothing to prove. Let $(\mu,\sigma)\in\K[[\x]]\rtimes\gotM$ and $(\mup,\sigmap)\in\Kxp\rtimes\gotMp$. We need to prove that $(\mu,\sigma)\rtimes(\mup,\sigmap)$ and
$(\mup,\sigmap)\rtimes(\mu,\sigma)$ both belong to $\Kxp\rtimes\gotMp$. On the one hand, the former product is equal to $((\mu\circ\sigmap)\mup,\sigma\circ\sigmap)$. Since we already know that $\sigma\circ\sigmap\in\Kxp$, we only need to establish that
$\nu((\mu\circ\sigmap)\mup)>0$. But $\nu((\mu\circ\sigmap)\mup)=\nu(\mu\circ\sigmap)+\nu(\mup)=\nu(\mu)\mu(\sigmap)+\nu(\mup)>0$. On the other hand, $(\mup,\sigmap)\rtimes(\mu,\sigma)=((\mup\circ\sigma)\mu,\sigmap\circ\sigma)$ and as in the first case, the only fact to check is $\nu((\mup\circ\sigma)\mu)>0$. But $\nu((\mup\circ\sigma)\mu)=\nu(\mup)\nu(\sigma)+\nu(\mu)>0$ (because both $\nu(\mup)$ and $\nu(\sigma)$ are positive).\\

Independently from algebraic considerations, it is possible to prove that each element of $\K[[\x]]$ operates on $\Kxp\rtimes\gotMp$.  The argument to prove this fact is partially based on the following lemma.

\begin{lemma}\label{operationseriesformellesurfaadibruno}
$\K[[\x]]$ operates on $\gotMp$. More precisely, for each $f=\displaystyle\sum_{n\geq 0}f_n \x^n \in \K[[\x]]$ and each
$\sigmap\in\gotMp$, $\displaystyle\sum_{n\geq 0}f_n \sigmap^{{\circ n}}\in\gotM$.
\end{lemma}
\begin{proof}
The goal to prove is the fact that for every $f=\displaystyle\sum_{n\geq 0}f_n\x^n\in\K[[\x]]$ and every $\sigmap\in\gotMp$, $\displaystyle\sum_{n\geq 0}f_n \sigmap^{{\circ n}}$ is a well-defined element of $\gotM$. According to lemma~\ref{ordredunecompositon}, $\nu(\sigmap^{{\circ n}})=\nu(\sigmap)^n\geq 2^n$ for every $n\in\N$. Therefore $\displaystyle\lim_{n\rightarrow+\infty}\nu(\sigmap^{{\circ n}})=+\infty$, so that the series $\displaystyle\sum_{n\geq 0}f_n \sigmap^{{\circ n}}$ converges in $\K[[\x]]$. Moreover it is easy to check that $\langle \displaystyle\sum_{n\geq 0}f_n \sigmap^{{\circ n}},1\rangle = 0$ and $\langle \displaystyle\sum_{n\geq 0}f_n\sigmap^{{\circ n}},\x\rangle=f_0$ (because $\sigmap^{\circ 0}=\x$). The convergence in $\gotM$ follows.
\end{proof}

\begin{proposition}\label{operation_series_formelle_sur_algebre}
$\K[[\x]]$ operates on $\Kxp\rtimes\gotMp$. More precisely for every $f=\displaystyle\sum_{n\geq 0}f_n\x^n\in\K[[\x]]$ and every $(\mup,\sigmap)\in\Kxp\rtimes\gotMp$, $\displaystyle\sum_{n\geq 0}f_n (\mup,\sigmap)^{\rtimes n}\in\K[[\x]]\rtimes\gotM$.
\end{proposition}
\begin{proof}
The goal to be proved is that $\displaystyle\sum_{n\geq 0}f_n (\mup,\sigmap)^{\rtimes n}$ is a convergent series in $\K[[\x]]\rtimes\gotM$ whenever $(\mup,\sigmap)\in\Kxp\rtimes\gotMp$. A proof by case follows.
\begin{enumerate}
\item $\mup=\sigmap=0$: For every $n>0$, $(0,0)^{\rtimes n}=(0,0)$. Therefore $\displaystyle\sum_{n\geq 0}f_n (0,0)^{\rtimes n}$ reduces to  $f_0(1,\x)\in\K[[\x]]\rtimes\gotM$;
\item $\mup=0$ and $\sigmap\not=0$: According to lemma~\ref{calculexplicitedelasommegeneraliseeduproduitsemidirect}, for every $n>0$, $(0,\sigmap)^{\rtimes n}=(0,\sigmap^{{\circ n}})$, so we only need to prove that the series $\displaystyle \sum_{n\geq 0}f_n \sigmap^{{\circ n}}$ is convergent in $\gotM$ which is the case by lemma~\ref{operationseriesformellesurfaadibruno} since $\sigmap\in\gotMp$;
\item $\mup\not=0$ and $\sigmap=0$: According to lemma~\ref{calculexplicitedelasommegeneraliseeduproduitsemidirect}, for every $n>0$, $(\mup,0)^{\rtimes n}=(\mup\mup(0)^{n-1},0)$, so we only need to prove that the series $$\displaystyle f_0+\sum_{n\geq 1}f_n \mup\mup(0)^{n-1}$$ converges in $\K[[\x]]$. Since $\mup\in \Kxp$, $\mup(0)=0$ so that $f_1\mup\mup(0)^0=f_1\mup$ and $f_n\mup\mup(0)^{n-1}=0$ for every $n>1$. Therefore $$\displaystyle f_0 + \sum_{n\geq 1}f_n \mup\mup(0)^n=f_0+f_1\mup\in\K[[\x]]\ .$$
\item $\mup\not=0$ and $\sigmap\not=0$: Using lemmas~\ref{calculexplicitedelasommegeneraliseeduproduitsemidirect} and \ref{operationseriesformellesurfaadibruno} it already holds that the second component of the series is convergent in $\gotM$ (since $\sigmap\in\gotMp$). Let us study the first component. For every $n>0$, taking into account lemma~\ref{calculexplicitedelasommegeneraliseeduproduitsemidirect}, $f_n(\mup,\sigmap)^{\rtimes n}=\displaystyle f_n(\prod_{k=1}^n(\mup\circ\sigmap^{\circ(k-1)}),\sigmap^{{\circ n}})$. We need to evaluate the valuation of $\displaystyle \prod_{k=1}^n(\mup\circ\sigmap^{\circ(k-1)})$: $\nu(\displaystyle \prod_{k=1}^n(\mup\circ\sigmap^{\circ(k-1)}))=\nu(\mup)\sum_{k=1}^n\nu(\sigmap)^{k-1}\geq \nu(\mup)\sum_{k=1}^n2^{k-1}$. Since $\nu(\mup)>0$, it follows that $\displaystyle\lim_{n\rightarrow+\infty}\nu(\prod_{k=1}^n (\mup\circ\sigmap^{\circ(k-1)}))=+\infty$ which ensures the convergence of the first component. Therefore the series $\displaystyle\sum_{n\geq 0}f_n (\mup,\sigmap)^{\rtimes n}$ is componentwise convergent and so is convergent in the product topology of $\K[[\x]]\rtimes\gotM$.
\end{enumerate}
\end{proof}

\begin{rem}
In the later proof, we implicitly show that every $(\mup,\sigmap)\in\Kxp\rtimes\gotMp$ is topologically nilpotent and even nilpotent in the usual sense when $\sigmap$ equals zero. Likewise, in the proof of lemma~\ref{operationseriesformellesurfaadibruno}, we also show that every $\sigmap\in\gotMp$ is topologically nilpotent in $\gotM$. 
\end{rem}

The above proposition guarantees the existence in $\K[[\x]]\rtimes\gotM$ of, for instance, $\exp(\mup,\sigmap)$ - defined as the sum of the series $\displaystyle\sum_{n\geq 0}\frac{1}{n!}(\mup,\sigmap)^{\rtimes n}$ - or $\displaystyle\sum_{n\geq 0}(\mup,\sigmap)^{\rtimes n}$ whenever $(\mup,\sigmap)\in\Kxp\rtimes\gotMp$. We can note that the later series generally does not define $((1,\x)-(\mup,\sigmap))^{\rtimes(-1)}$ as we would expect since in general $((1,\x)-(\mup,\sigmap))\rtimes\displaystyle\left ( \sum_{n=0}^m(\mup,\sigmap)^{\rtimes n}\right )\not=\displaystyle\left (\sum_{n=0}^m (\mup,\sigmap)^{\rtimes n} \right ) \rtimes ((1,\x)-(\mup,\sigmap))$ because of noncommutativity of $\rtimes$ and its lack of left distributivity. Nevertheless it will soon be shown (see section~\ref{sect4}) that $\displaystyle\sum_{n\geq 0}(\mup,\sigmap)^{\rtimes n}$ is the inverse of $((1,\x)-(\mup,\sigmap))$ for another kind of multiplication. \\

As another direct consequence of the above proposition, we have the following result.
Let $(\mu,\sigma)\in \K[[\x]]\rtimes\gotM$ and $f=\displaystyle\sum_{n\geq 0}f_n\x^n\in\K[[\x]]$. Then $\displaystyle\sum_{n\geq 0}f_n ((\mu,\sigma)-(\mu(0),\langle \sigma,\x\rangle\x))^{\rtimes n}\in \K[[\x]]\rtimes \gotM$. This result is indeed straightforward because $(\mu,\sigma)-(\mu(0),\langle \sigma,\x\rangle \x)\in \Kxp\rtimes \gotMp$ whenever $(\mu,\sigma)\in\K[[\x]]\rtimes\gotM$. \\

The operation of $\K[[\x]]$ on $\Kxp\rtimes\gotMp$ gives rise to the following mapping.
\begin{equation}
\begin{array}{llll}
\Psi: & \K[[\x]]\times(\Kxp\rtimes\gotMp) & \rightarrow & \K[[\x]]\rtimes\gotM\\
& (f,(\mup,\sigmap)) & \mapsto & \displaystyle\sum_{n\geq 0}f_n (\mup,\sigmap)^{\rtimes n}
\end{array}
\end{equation}
where $f=\displaystyle\sum_{n\geq 0}f_n\x^n$. This operation has some interesting properties stated below, even if they are not important for the main subject of the paper.

\begin{lemma}
Let $(\mup,\sigmap)\in\Kxp\rtimes\gotMp$. We define
\begin{equation}
\begin{array}{llll}
\phi_{(\mup,\sigmap)}: & \K[[\x]] & \rightarrow & \K[[\x]]\rtimes\gotM\\
& f & \mapsto & \Psi(f,(\mup,\sigmap))\ .
\end{array}
\end{equation}
Then, $\phi_{(\mup,\sigmap)}$ is a vector space homomorphism that maps $\x$ to $(\mup,\sigmap)$. Moreover if $f\in\gotM$, then $\phi_{(\mup,\sigmap)}(f)\in\Kxp\rtimes\gotMp$, and if $\langle f,1\rangle=1$, then $\phi_{(\mup,\sigmap)}(f)\in \UP\rtimes\US$.
\end{lemma}
\begin{proof}
Let $f=\displaystyle\sum_{n\geq 0}f_n \x^n$ and
$g=\displaystyle\sum_{n\geq 0}g_n \x^n$ be two formal series. We have $f+g=\displaystyle\sum_{n\geq 0}(f_n+g_n)\x^n$. Besides  $\phi_{(\mup,\sigmap)}(f)=\displaystyle\sum_{n\geq 0}f_n (\mup,\sigmap)^{\rtimes n}$ and $\phi_{(\mup,\sigmap)}(g)=\displaystyle\sum_{n\geq 0}g_n (\mup,\sigmap)^{\rtimes n}$ and finally  $\phi_{(\mup,\sigmap)}(f+g)=\displaystyle\sum_{n\geq 0}(f_n+g_n)(\mup,\sigmap)^{\rtimes n}=\sum_{n\geq 0}(f_n (\mup,\sigmap)^{\rtimes n}+g_n(\mup,\sigmap)^{\rtimes n})=\sum_{n\geq 0}f_n(\mup,\sigmap)^{\rtimes n}+\sum_{n\geq 0}g_n(\mup,\sigmap)^{\rtimes n}$ (the last equality is due to the fact that $\K[[\x]]\rtimes \gotM$ is a topological group). 
Scalar multiplication by $\alpha \in \K$ is continuous on $\K[[\x]]\rtimes \gotM$ and one has $\phi_{(\mup,\sigmap)}(\alpha f)=\displaystyle\sum_{n\geq 0}\alpha f_n (\mup,\sigmap)^{\rtimes n}=\alpha \sum_{n\geq 0}f_n (\mup,\sigmap)^{\rtimes n}=\alpha \phi_{(\mup,\sigmap)}(f)$. Finally the last statements are  rather straightforward. 
\end{proof}

In order to deeply study $\phi_{(\mup,\sigmap)}$ another easy lemma is needed.
\begin{lemma}
Let $(\mup,\sigmap)\in\Kxp\rtimes \gotMp$, $g\in \K[[\x]]$ and $m\in\N$. Then
\begin{equation}
\phi_{(\mup,\sigmap)}(\x^m g)=\phi_{(\mup,\sigmap)}(g)\rtimes (\mup,\sigmap)^{\rtimes m}=\phi_{(\mup,\sigmap)}(g\x^m)\ .
\end{equation}
\end{lemma}
\begin{proof}
\begin{equation}
\begin{array}{lll}
\phi_{(\mup,\sigmap)}(\x^m g)&=&\phi_{(\mup,\sigmap)}(\displaystyle\sum_{n=0}^{\infty}g_n \x^{n+m})\\
&=&\displaystyle\sum_{n\geq 0}g_n (\mup,\sigmap)^{\rtimes(n+m)}\\
&=&\displaystyle\sum_{n\geq 0}g_n\left ( (\mup,\sigmap)^{\rtimes n}\rtimes (\mup,\sigmap)^{\rtimes m}\right )\\
&=&\displaystyle\sum_{n\geq 0}\left( (g_n (\mup,\sigmap)^{\rtimes n})\rtimes (\mup,\sigmap)^{\rtimes m }\right )\\
&&(\mbox{according to the rule of right distributivity}.)\\
&=&\displaystyle\left ( \sum_{n\geq 0}g_n (\mup,\sigmap)^{\rtimes n}\right ) \rtimes (\mup,\sigmap)^{\rtimes m}\\
&&(\mbox{by continuity and linearity of $\rtimes$ in its first variable}.)\\
&=&\phi_{(\mup,\sigmap)}(g)\rtimes (\mup,\sigmap)^{\rtimes m}\ .
\end{array}
\end{equation}
\end{proof}


\begin{proposition}
Let $(\mup,\sigmap)\in\Kxp\rtimes \gotMp$. Then $\phi_{(\mup,\sigmap)}$ is the only linear mapping $\psi : \K[[\x]] \rightarrow \K[[\x]]\rtimes \gotM$ such that for every $m\in \N$ and every $g\in \K[[\x]]$, $\psi(g\x^m)=\psi(g)\rtimes (\mup,\sigmap)^{\rtimes m}$.
\end{proposition}
\begin{proof}
Let $f \in \K[[\x]]$. For every $m\in\N$, $f=\displaystyle\sum_{n=0}^m f_n \x^n + \x^{m+1}g$ with $g \in \K[[\x]]$.
Let $\psi$ as in the statement of the proposition. One has $\psi(f)=\displaystyle\sum_{n=0}^m f_n (\mup,\sigmap)^{\rtimes n}+\psi(g)\rtimes(\mup,\sigmap)^{\rtimes (m+1)}$ and similarly, the following equality also holds $\phi_{(\mup,\sigmap)}(f)=\displaystyle\sum_{n=0}^m f_n (\mup,\sigmap)^{\rtimes n}+\phi_{(\mup,\sigmap)}(g)\rtimes(\mup,\sigmap)^{\rtimes (m+1)}$. Then it follows that
$\psi(f)-\phi_{(\mup,\sigmap)}(f)=(\psi(g)-\phi_{(\mup,\sigmap)}(g))\rtimes (\mup,\sigmap)^{\rtimes (m+1)}$ for every $m\in\N$. But when $m\rightarrow +\infty$, $(\mup,\sigmap)^{\rtimes (m+1)}$ converges to $(0,0)$. Indeed, suppose that $\mup=0$ and $\sigmap=0$, then the result obviously holds. If $\mup=0$ and $\sigmap\not=0$, then $(0,\sigmap)^{\rtimes (m+1)}=(0,\sigmap^{\circ (m+1)})$ and   $\nu(\sigmap^{\circ (m+1)})=v(\sigmap)^{m+1}\geq 2^{m+1}$. If $\mup\not=0$ and $\sigmap=0$, then  $(\mup,0)^{\rtimes (m+1)}=(\mup\mup(0)^m,0)$. Since $\mup(0)=0$ (because $\mup\in\Kxp$), for every $m>0$, $(\mup\mup(0)^m,0)=(0,0)$. Finally let suppose that $\mup\not=0$ and $\sigmap\not=0$. Therefore
$$(\mup,\sigmap)^{\rtimes (m+1)}=(\displaystyle\prod_{k=1}^{m+1} (\mup \circ \sigmap^{\circ(k-1)}),\sigmap^{\circ (m+1)})\ .$$ We already know that $\displaystyle\lim_{m\rightarrow\infty}\sigmap^{\circ(m+1)}=0$. We also have $\nu(\displaystyle\prod_{k=1}^{m+1} (\mup \circ \sigmap^{\circ(k-1)}))=
\sum_{k=1}^{m+1}\nu(\mup)\nu(\sigmap)^{\circ(k-1)}\geq \nu(\sigma)\sum_{k=1}^{m+1}2^{k-1}\rightarrow \infty$ as $m\rightarrow \infty$.  Besides we have seen in lemma~\ref{continuitesepareedelamultiplication} that for every $a \in \K[[\x]]\rtimes \gotM$, the mapping
\begin{equation}
\begin{array}{llll}
L_a: & \K[[\x]]\rtimes \gotM & \rightarrow &\K[[\x]]\rtimes \gotM\\
& (\mu,\sigma) & \mapsto & a\rtimes (\mu,\sigma)
\end{array}
\end{equation}
is continuous and in particular at the point $(0,0)$. Since the topology put on $\K[[\x]]\rtimes \gotM$ is metrizable (as the product of two metric topologies), then for every sequence  $(b_n)_n \in (\K[[\x]]\rtimes\gotM)^{\N}$ converging to $(0,0)$, one has  $\displaystyle \lim_{n\rightarrow\infty}L_a(b_n)=L_a(0,0)=(0,0)$. When applied to the case  $a:=(\psi(g)-\phi_{(\mup,\sigmap)}(g))$ and  $b_n=(\mup,\sigmap)^{\rtimes (n+1)}$, we deduce that $\psi(f)=\phi_{(\mup,\sigmap)}(f)$ for an arbitrary formal power series $f$, so $\psi=\phi_{(\mup,\sigmap)}$.
\end{proof}

\subsection{Generalized powers by binomial series}

In this subsection is presented a result which seems to provide a relevant definition for generalized powers of elements of the Riordan group. However we will be shown that it is not at all the case, and we will have to propose another solution in the subsequent section. Recall that we have
\begin{equation}
\begin{array}{lllll}
\UP &:=& \{1+\x s : s\in\K[[\x]]\}&=&\{1+\mup : \mup\in \K[[\x]]^{+}\}\ ,\\
\US &:=& \{\x + \x^2 s : s\in \K[[\x]]\}&=&\{\x+\sigmap : \sigmap\in\gotMp\}\ .
\end{array}
\end{equation}
The elements of $\mathsf{US}$ are also known under the name ``formal diffeomorphisms (tangent to the identity)" (see for instance~\cite{BFK06}). The semi-direct product $\UP\rtimes \US$, called ``Riordan group" (\cite{SGW$^+$91}), is a subgroup of the group of units of $\K[[\x]]\rtimes\gotM$. It endows the subspace topology as usually. \\

We now recall the traditional definition for generalized binomial coefficients: let $\lambda \in \K$ and $n\in \N$, then
$\left( {\begin{array}{*{20}c}
\lambda\\
n\\
\end{array}}\right):=\frac{\lambda (\lambda -1)\dots (\lambda - n+1)}{n!}$.
Now let us prove a statement similar to proposition~4.1~\cite{DPS$^+$04} in our setting.
\begin{proposition}\label{c_presque_bon_mais_pas_encore}
Let $(\mu,\sigma)\in \UP\rtimes \US$ with $\mu=1+\mup$, $\mup\in\Kxp$ and $\sigma=\x+\sigmap$, $\sigmap\in\gotMp$. Let $\lambda \in \K$. Then the series
$(\mu,\sigma)^{\rtimes \lambda}=((1,\x)+(\mup,\sigmap))^{\rtimes \lambda}:=\displaystyle \sum_{n\geq 0}\left( {\begin{array}{*{20}c}
\lambda\\
n\\
\end{array}}\right) (\mup,\sigmap)^{\rtimes n}$ is convergent in $\K[[\x]]\rtimes\gotM$ and the sum of this series belongs to $\UP\rtimes\US$.
\end{proposition}

\begin{proof}
According to proposition~\ref{operation_series_formelle_sur_algebre} we already agree for the convergence of the series in $\K[[\x]]\rtimes\gotM$. To conclude the proof it is sufficient to check that the sum of the series belongs to $\UP\rtimes\US$. The first term of the series is $(1,\x)$ because $\left( {\begin{array}{*{20}c}
\lambda\\
0\\
\end{array}}\right)=1$. Now we make use of lemma~\ref{calculexplicitedelasommegeneraliseeduproduitsemidirect} to study the terms $(\mup,\sigmap)^{\rtimes n}$ for each $n\in\N\setminus\{0\}$.
\begin{enumerate}
\item Second coordinate of $(\mup,\sigmap)^{\rtimes n}$:
\begin{itemize}
\item Case $\sigmap=0$: the second component is equal to $0$ for every $n\in\N\setminus\{0\}$;
\item Case $\sigmap\not=0$: the second component is equal to $\sigmap^{{\circ n}}$. According to lemma~\ref{ordredunecompositon}, $\nu(\sigma^{{\circ n}})\geq 2^n>1$.
\end{itemize}
\item First coordinate of $(\mup,\sigmap)^{\rtimes n}$:
\begin{itemize}
\item Case $\mup=0$: the first component is equal to $0$ for every $n\in\N\setminus\{0\}$;
\item Case $\mup\not=0$:
\begin{itemize}
\item Case $\sigmap=0$: the first component is equal to $\mup\mup(0)^{n-1}=\mup$ if $n=1$ and to $0$ if $n>1$ since $\mup(0)=0$;
\item Case $\sigmap\not=0$: the first component is equal to $\displaystyle\prod_{k=1}^n \mup\circ\sigmap^{\circ(k-1)}$. Then according to lemma~\ref{ordredunecompositon}, $$\nu(\displaystyle\prod_{k=1}^n \mup\circ\sigmap^{\circ(k-1)})\geq \nu(\mup)\sum_{k=1}^n 2^{k-1}>0\ .$$
\end{itemize}
\end{itemize}
\end{enumerate}
\end{proof}

The definition of generalized powers for elements of the Riordan group provided by the previous proposition seems quite natural,  nevertheless it is not the convenient one in our setting. Actually when restricted to natural integers it does not match with the usual powers in $\K[[\x]]\rtimes\gotM$ as it can be easily checked even on trivial instances: let $\mu=1+\x$ and $\sigma=\x+\x^2$. Therefore on the one hand, seen as an element of the Riordan group, one has
\begin{equation}
\begin{array}{lll}
(\mu,\sigma)^{\rtimes 2}&=&((\mu\circ \sigma)\times\mu,\sigma^{\circ 2})\\
&=&(((1+\x)\circ(\x+\x^2))\times (1+\x),(\x+\x^2)\circ(\x+\x^2))\\
&=&((1+\x+\x^2)(1+\x),\x+\x^2 +(\x+\x^2)^2)\\
&=&(1+2\x+2\x^2+\x^3,\x+2\x^2+2\x^3+\x^4)\ .
\end{array}
\end{equation}
Using the series definition, we have on the other hand,
\begin{equation}
\begin{array}{lll}
(\x,\x^2)^{\rtimes 0} + 2(\x,\x^2)^{\rtimes 1}+(\x,\x^2)^{\rtimes 2}&=&(1,\x)+2(\x,\x^2)+(\x,\x^2)\rtimes(\x,\x^2)\\
&=&(1,\x)+(2\x,2\x^2)+(\x^3,\x^4)\\
&=&(1+2\x+\x^3,\x+2\x^2+\x^4)\ .
\end{array}
\end{equation}

So our definition for generalized powers has a serious weakness: it does not generalize the usual powers, which makes it impossible to be taken as generalized powers at least in this minimal sense. The same weakness is shared by the exponential, logarithm or inverse series for instance. Nevertheless there is a convenient algebra in which those series play their expected roles.

\section{A convenient setting for the generalized powers}\label{sect4}

\subsection{The algebra of formal power series $\K[[\mup,\sigmap]]$}

In order to fix the problem met in the end of the previous section, we need to introduce a new algebra in which can be lead convenient calculus.\\
Let $(\mup,\sigmap)\in\Kxp\rtimes \gotMp\setminus\{(0,0)\}$. Our first goal is to prove that $\phi_{(\mup,\sigmap)}$ is one-to-one.
\begin{lemma}
For every integers $n<m$,
\begin{enumerate}
\item $\displaystyle \nu(\prod_{k=1}^n\mup\circ\sigmap^{\circ(k-1)})<\nu(\prod_{k=1}^n\mup\circ\sigmap^{\circ(k-1)})$;
\item $\displaystyle \nu(\sigmap^{{\circ n}})<\nu(\sigmap^{\circ m})$.
\end{enumerate}
\end{lemma}
\begin{proof}
\begin{enumerate}
\item
\begin{itemize}
\item Suppose that $n=0$ (and therefore $m>0$). In this case $\displaystyle\prod_{k=1}^0\mup\circ\sigmap^{\circ(k-1)}:=1$ by convention and then $\nu(1)=0$. Besides $\displaystyle \nu(\prod_{k=1}^n\mup\circ\sigmap^{\circ(k-1)})=\nu(\mup)\sum_{k=1}^m \nu(\sigmap^{k-1})>0$;
\item Suppose that  $n> 0$. Then it is clear that the choices of $\mup$ and $\sigmap$  gives the expected result.
\end{itemize}
\item
\begin{itemize}
\item Suppose that $n=0$. $\nu(\sigmap^{\circ 0})=\nu(\x)=1$ and $\nu(\sigmap^{\circ m})=\nu(\sigmap)^m\geq 2^m>1$ for every $m>0$;
\item Suppose that $n\not=0$. Then it is clear that $\nu(\sigmap)^n<\nu(\sigmap)^m$.
\end{itemize}
\end{enumerate}
\end{proof}

\begin{lemma}
$\phi_{(\mup,\sigmap)}:\K[[\x]] \rightarrow \K[[\x]]\rtimes\gotM$ is one-to-one.
\end{lemma}
\begin{proof}
Since it is a linear mapping, it is sufficient to check that its kernel is reduced to zero. So let  $f=\displaystyle\sum_{n=0}^{\infty}f_n\x^n\in\K[[\x]]$ be a nonzero series. Let $n_0:=\nu(f)$. In this case,  $\phi_{(\mup,\sigmap)}(f)=\phi_{(\mup,\sigmap)}(f_{n_0}\x^{n_0}+\displaystyle\sum_{n>n_0}f_n \x^n)=
\phi_{(\mup,\sigmap)}(f_{n_0}\x^{n_0})+\phi_{(\mup,\sigmap)}(\sum_{n>n_0}f_n \x^n)=f_{n_0}(\mup,\sigmap)^{\rtimes n_0}+
\displaystyle\sum_{n>n_0}f_n (\mup,\sigmap)^{\rtimes n}$. Checking component by component and using the previous lemma, we obtain the expected result.
\end{proof}

\begin{rem}
The mapping $\phi_{(\mup,\sigmap)}$ is far from being onto. For instance, let $(\mu,\sigma)\in \K[[\x]]\rtimes\gotM$ such that $\mu=\alpha+\mu_{+}$ and
$\sigma=\beta\x + \sigma_{+}$ with $\alpha \not=\beta$, $\alpha\not=0$ and $\beta\not=0$. If $(\mu,\sigma)\in\K[[\mup,\sigmap]]$ then there exists $f\in\K[[\x]]$ such that $(\mu,\sigma)=\displaystyle\sum_{n\geq 0}f_n(\mup,\sigmap)^{\rtimes n}$ and, according to the fact that $(\mup,\sigmap)\in \Kxp\rtimes\gotMp$, $f_0(1,\x)=(\alpha,\beta\x)$, which implies that $f_0=\alpha=\beta$: a contradiction.
\end{rem}

Now it becomes natural to define $$\K[[\mup,\sigmap]]:=\im(\phi_{\mup,\sigmap})=\{\displaystyle \sum_{n\geq 0}f_n (\mup,\sigmap)^{\rtimes n} : f=\sum_{n\geq 0}f_n \x^n \in \K[[\x]]\}\ .$$ By injectivity of $\phi_{(\mup,\sigmap)}$, for every $(\mu,\sigma)\in \K[[\mup,\sigmap]]$, it exists one only one formal power series $f$ such that  $(\mu,\sigma)=\displaystyle\sum_{n\geq 0}f_n (\mup,\sigmap)^{\rtimes n}$. So it is possible to manipulate the elements of  $\K[[\mup,\sigmap]]$ \emph{via} their representation as a sum of converging series in the ``variable" $(\mup,\sigmap)$. It is also interesting to remark, due to lemma~\ref{topological_nilpotence_in_dibruno} since $(\mup,\sigmap)$ is topologically nilpotent, that $\displaystyle \sum_{n\in\N}f_n (\mup,\sigmap)^{\rtimes n}$ does not depend on the order of summation (which explains the use of the notation ``$n\in\N$" rather than ``$n\geq 0$"). Because $\phi_{(\mup,\sigmap)}$ is a linear mapping, $\K[[\mup,\sigmap]]$ has a structure of $\K$ subvector space of $\K[[\x]]\rtimes \gotM$. In particular,
$\displaystyle\lambda \left ( \sum_{n\geq 0}f_n (\mup,\sigmap)^{\rtimes n}\right )=\displaystyle\sum_{n\geq 0}\lambda f_n (\mup,\sigmap)^{\rtimes n}$ and
$\displaystyle \sum_{n\geq 0}f_n (\mup,\sigmap)^{\rtimes n}+\sum_{n\geq 0}g_n (\mup,\sigmap)^{\rtimes n}=\sum_{n\geq 0}(f_n + g_n)(\mup,\sigmap)^{\rtimes n}$. The addition of two elements of $\K[[\mup,\sigmap]]$ in $\K[[\x]]\rtimes\gotM$ matches with their addition in $\K[[\mup,\sigmap]]$.  Nevertheless the notation  $\K[[\mu_{+},\sigma_{+}]]$ should seem misleading because $\phi_{(\mup,\sigmap)}(fg)\not=\phi_{(\mup,\sigmap)}(f)\rtimes \phi_{(\mup,\sigmap)}(g)$. Indeed, on the one side,
\begin{equation}
\begin{array}{lll}
\displaystyle\phi_{(\mup,\sigmap)}(fg)&=&\displaystyle\sum_{n\geq 0}\left (\sum_{k=0}^n f_k g_{n-k}\right )(\mup,\sigmap)^{\rtimes n}
\end{array}
\end{equation}
and on the other side,
\begin{equation}
\begin{array}{lll}
\displaystyle\phi_{(\mup,\sigmap)}(f)\rtimes\phi_{(\mup,\sigmap)}(g)&=&\displaystyle\left ( \sum_{n\geq 0}f_n(\mup,\sigmap)^{\rtimes n}\right ) \rtimes \left( \sum_{n\geq 0}g_n(\mup,\sigmap)^{\rtimes n}\right )\\
&=&\displaystyle\sum_{n\geq 0}f_n \left ((\mup,\sigmap)^{\rtimes n} \rtimes\left ( \sum_{k\geq 0}g_k (\mup,\sigmap)^{\rtimes k}\right)\right )\\
&&(\mbox{by linearity and continuity in the first variable of $\rtimes$}.) \\
&=&\displaystyle\sum_{n\geq 0}f_n \left ( (\mup,\sigmap)\rtimes \phi_{(\mup,\sigmap)}(g)\right )
\end{array}
\end{equation}
In order to obtain an algebra, we introduce the usual Cauchy product ``$*$" on $\K[[\mup,\sigmap]]$.
\begin{equation}
\displaystyle\sum_{n\geq 0}f_n (\mup,\sigmap)^{\rtimes n}*\sum_{n\geq 0}g_n (\mup,\sigmap)^{\rtimes n}:=\sum_{n\geq 0}\left ( \sum_{k=0}^{n}f_k g_{n-k}\right )(\mup,\sigmap)^{\rtimes n}\ .
\end{equation}
We should remark that this multiplication is commutative contrary to $\rtimes$. Actually this operation simulates the multiplication $\rtimes$ in the multiplicative monoid generated by $(\mup,\sigmap)$. Indeed let $d\in\N$. We define  $(\delta_n^{(d)})_{n\in\N}\in\K^{\N}$ by $\delta_n^{(d)}=0$ for every $n\not=d$ and $\delta_d^{(d)}=1$. Then we have  $\displaystyle\sum_{n\geq 0}\delta_n^{(d)}(\mup,\sigmap)^{\rtimes n}=(\mup,\sigmap)^{\rtimes d}$. Now let $d,e\in\N$. Let us compute the following Cauchy product 
\begin{equation}
\begin{array}{lll}\displaystyle\sum_{n\geq 0}\delta_n^{(d)}(\mup,\sigmap)^{\rtimes n}*\sum_{n\geq 0}\delta_n^{(e)}(\mup,\sigmap)^{\rtimes n}&=&\displaystyle 
\sum_{n\geq 0}\left (\underbrace{\sum_{k=0}^n \delta_k^{(d)}\delta_{n-k}^{(e)}}_{=0\ \Leftrightarrow\ k\not=d,n\not=d+e}\right )(\mup,\sigmap)^{\rtimes n}\\
&=&(\mup,\sigmap)^{\rtimes (d+e)}\ .
\end{array}
\end{equation}
But the first member of the Cauchy product occurring as the left member of the first equality is nothing else than $(\mup,\sigmap)^{\rtimes d}$, whereas its second member is $(\mup,\sigmap)^{\rtimes e}$. On ``monomials" $(\mup,\sigmap)^{\rtimes n}$ the Cauchy product is identical to $\rtimes$. In particular for every natural integer $n$, $(\mup,\sigmap)^{\rtimes n}=(\mup,\sigmap)^{*n}$ where the second member is the $n$th Cauchy power of
$(\mup,\sigmap)$. Then, $\phi_{(\mup,\sigmap)}$ becomes an algebra isomorphism from $\K[[\x]]$ into $\K[[\mup,\sigmap]]$. \\

We use this Cauchy product to define the generalized power of elements $(1+\mup,\x+\sigmap)$ of the Riordan group in terms of a the binomial series: let $\lambda \in \K$ and define $\displaystyle ((1,\x)+(\mup,\sigmap))^{*\lambda}:=\displaystyle \sum_{n\geq 0}\left( {\begin{array}{*{20}c}
\lambda\\
n\\
\end{array}}\right) (\mup,\sigmap)^{*n}$. We need to prove that this binomial series is convergent. Nevertheless it can be checked that if $\lambda\in \N$, then $((1,\x)+(\mup,\sigmap))^{*\lambda}$ matches with the $\lambda$th Cauchy power of $(1+\mup,\x+\sigmap)\in\K[[\mup,\sigmap]]$. Therefore this version of the generalized powers extends the usual ones (in $\K[[\mup,\sigmap]]$ not in $\UP\rtimes\US$).

\begin{proposition}
The series $\displaystyle ((1,\x)+(\mup,\sigmap))^{*\lambda}=\displaystyle \sum_{n\geq 0}\left( {\begin{array}{*{20}c}
\lambda\\
n\\
\end{array}}\right) (\mup,\sigmap)^{*n}$ is convergent and defines an element of $\UP\rtimes\US$.
\end{proposition}
\begin{proof}
Actually since $(\mup,\sigmap)^{*n}=(\mup,\sigmap)^{\rtimes n}$, the result is already given by proposition~\ref{c_presque_bon_mais_pas_encore}.
\end{proof}

Mimicking in $\K[[\mup,\sigmap]]$ the usual properties that hold true in the algebra $\K[[\x]]$ of formal power series\footnote{To be more rigorous, we need to equip $\K[[\mup,\sigmap]]$ with the $(\mup,\sigmap)$-adic topology or, equivalently, with the valuation obtained from $\K[[\x]]$'s one by replacing the monomials in $\x$ by monomials in $(\mup,\sigmap)$: in other terms, one can use $\phi_{(\mup,\sigmap)}$ to transport the topology of $\K[[\x]]$ on $\K[[\mup,\sigmap]]$ in a homeomorphic way. Then $\K[[\mup,\sigmap]]$ becomes isomorphic to $\K[[\x]]$ as a topological algebra and $\phi_{(\mup,\sigmap)}$ becomes a topological isomorphism.}, we can check that for every $\lambda \in \K$,
\begin{equation}
\begin{array}{lll}
\exp(\lambda \log((1,\x)+(\mup,\sigmap)))=((1,\x)+(\mup,\sigmap))^{*\lambda}\ .
\end{array}
\end{equation}
Moreover, due to the fact that for every $f\in \K[[\x]]$, $\exp(\displaystyle\sum_{n\geq 0}f_n(\mup,\sigmap)^{\rtimes n})$ is invertible in $\K[[\mup,\sigmap]]$, $\lambda \mapsto ((1+\mup,\x+\sigmap))^{*\lambda}$ is easily seen as a one-parameter subgroup from $(\K,+,0)$ to $U(\K[[\mup,\sigmap]])$. \\

In summary this new version for generalized powers satisfies,
\begin{enumerate}
\item whenever $n\in\N$, $(1+\mup,\x+\sigmap)^{*n}$ is the usual $n$th power (with respect to Cauchy product) of $(1+\mup,\x+\sigmap)$ as an element of $\K[[\mup,\sigmap]]$ but not as an element of $\UP\rtimes \US$;
\item $(1+\mup,\x+\sigmap)^{*(-1)}$ is the inverse of $(1+\mup,\x+\sigmap)$ as an invertible element of $\K[[\mup,\sigmap]]$ but not as an element of $\UP\rtimes\US$;
\item $\lambda \mapsto (1+\mup,\x+\sigmap)^{*\lambda}$ is a one-parameter subgroup from $(\K,+,0)$ to $U(\K[[\mup,\sigmap]])$.
\end{enumerate}
In this setting, the following also holds. If $f_0=1$, then $\displaystyle\sum_{n\geq 0}f_n(\mup,\sigmap)^{\rtimes n}$ is invertible (in $\K[[\mup,\sigmap]]$) and its inverse is $\displaystyle\left ( \sum_{n\geq 0}f_n(\mup,\sigmap)^{\rtimes n}\right )^{*(-1)}=\sum_{n\geq 0}(-1)^n g(\mup,\sigmap)^{*n}$ where $g(\mup,\sigmap):=\displaystyle\sum_{n\geq 1}f_n(\mup,\sigmap)^{\rtimes n}$. For such a series $\displaystyle\sum_{n\geq 0}f_n(\mup,\sigmap)^{\rtimes n}$ which belongs to $\UP\rtimes \US$, we can also define for every $\lambda\in \K$, $\displaystyle\left (\sum_{n\geq 0}f_n (\mup,\sigmap)^{\rtimes n}\right )^{*\lambda}:=\sum_{n\geq 0}\left( {\begin{array}{*{20}c}
\lambda\\
n\\
\end{array}}\right)(g(\mup,\sigmap))^{*n}$ with the usual properties of one-parameter group of such generalized powers.

\subsection{An infinite number of copies of $\K[[\x]]$}

Because $\K[[\mup,\sigmap]]$ is isomorphic, as an algebra, to $\K[[\x]]$, it is possible to study the properties of series in powers of $(\mup,\sigmap)$ through the properties of the corresponding formal power series. We denote $\phi_{(\mup,\sigmap)}(f)$ by $f(\mup,\sigmap)$, and $\nu(f(\mup,\sigmap)):=\nu(f)$. Now $\K[[\mup,\sigmap]]$ is a isomorphic as a topological algebra to $\K[[\x]]$. In particular if $f=1+g \in\K[[\x]]$ where $\nu(g)>0$, then $f(\mup,\sigmap)$ has a multiplicative inverse in $\K[[\mup,\sigmap]]$ given by $\displaystyle\sum_{n\geq 0}(-1)^ng(\mup,\sigmap)^{*n}$, as already computed in  the previous subsection. Moreover if $\sigma\in\gotM$, then right substitution by $\sigma(\mup,\sigmap)$ is valid in $\K[[\mup,\sigmap]]$: $f(\mup,\sigmap)\circ \sigma(\mup,\sigmap):=\displaystyle\sum_{n\geq 0}f_n \sigma(\mup,\sigmap)^{*n}\in\K[[\mup,\sigmap]]$. So using $\K[[\mup,\sigmap]]$ we may define a near algebra  $\K[[\mup,\sigmap]]\rtimes\gotM(\mup,\sigmap)$ - where $\gotM(\mup,\sigmap):=(\mup,\sigmap)\rtimes\K[[\mup,\sigmap]]$ - isomorphic (both as a vector space and as a monoid) - to $\K[[\x]]\rtimes\gotM$. If $(\mu(\mup,\sigmap)_{+},\sigma(\mup,\sigmap)_{+})$ belongs to  $\K[[\mup,\sigmap]]^{+}\rtimes\gotM^{+}(\mup,\sigmap)$ (for the natural definitions of both $\K[[\mup,\sigmap]]^{+}$ and $\gotMp(\mup,\sigmap)$), then an algebra $\K[[\mu(\mup,\sigmap)_{+},\sigma(\mup,\sigmap)_{+}]]$, isomorphic to $\K[[\x]]$, may be defined. The process can continue indefinitely.
\begin{rem}
The set $\gotM(\mup,\sigmap)$, as $\gotM $, is both a near algebra and a ringoid. 
\end{rem}

\end{document}